\documentclass[twocolumn,preprintnumbers,amsmath,amssymb]{revtex4}
\usepackage{graphicx}

\usepackage{amsmath}
\usepackage{amssymb}
\usepackage{amsthm}
\usepackage{amsfonts}
\usepackage{enumerate}

\usepackage{centernot}
\usepackage{tikz}

\begin{document}

\title{Quantum chaos associated with emergent {ergosurface}  in transition layer between type-I and type-II Weyl semimetals }

\author{Yu-Ge Chen$^{1,2,3}$}
 \thanks{These two authors contribute equally.} 
\author{Xi Luo$^4$}
\thanks{These two authors contribute equally.}
\author{Fei-Ye Li$^{1,2,3}$}
\author{ Bin Chen$^4$}
\author{ Yue Yu$^{1,2,3}$}

\affiliation {
1. State Key Laboratory of Surface Physics, Fudan University, Shanghai 200433,
China\\
2.Center for Field
Theory and Particle Physics, Department of Physics, Fudan University, Shanghai 200433,
China\\
 3. Collaborative Innovation Center of Advanced Microstructures, Nanjing 210093, China\\
4. College of Science, University of Shanghai for Science and Technology, Shanghai 200093, PR China}

\date{\today}

\begin{abstract}

We present emergent ergosurfaces (ES) in a transition layer between type-I and type-II Weyl semimetals (WSMs).  The Hawking temperature defined by the surface gravity at the acoustic event horizon which coincides with the ES when the tangent velocity $v_{\parallel}$ is small is in a measurable interval.  On the type-II WSM side, i.e., inside the {ES when $v_{\parallel}$ is large}, the motion of the quasiparticles may be chaotic {after} a critical surface as they are governed by an effective inverted oscillator potential induced by the mismatch between the type-I and type-II Weyl nodes. In a relevant lattice model, we calculate out of time ordered correlators (OTOCs).  We find that the OTOCs are fast scrambling with a quantum Lyapunov exponent in high temperature and the scrambling is saturated after the Ehrenfest time. This confirms the quantum chaotic behavior.    \end{abstract}

\pacs{}

\maketitle

\section{ Introductions} Recent developments on the frontier of theoretical physics research have entangled the black hole horizon physics with quantum chaos \cite{mald2,kit1,kit2,mald1,sac,she1,she2}.

 In condensed matter systems, the black hole analog had been pioneered by Unruh in transsonic fluid flow \cite{unruh1}. A quantum analog had been provided in Bose-Einstein condensation \cite{BEC-BH,N19}, and a magnonic black hole was predicted \cite{molin}. The exotic emergent geometries  (or gravities) have also been studied in Fermi surface \cite{Horava}, fractional quantum Hall effects \cite{hald,qiu,ky,luo}, graphene \cite{vo6}, deformed crystal \cite{niu}, type-II Weyl semimetal (WSM) \cite{vo3,vo1,vo2,vo5}, and type-III Dirac semimetal \cite{liu}. Several systems may simulate the Hawking evaporation at a black hole event horizon \cite{hawkin,beken}. However, the chaotic behaviors at the horizon do not appear easily. For instance, for the trajectory of a moving particle outside of the horizon to be chaotic classically, there must be an external potential near the horizon \cite{hash}. The Lyapunov exponent is subject to a maximal chaotic bound, the surface gravity \cite{mald2}. This bound is saturated for theories with AdS/CFT duality, such as the case in the Sachdev-Ye-Kitaev model \cite{mald1,kit1,kit2}.

Volovik et al recently studied the emergent metric in the inhomogeneous WSM \cite{vo1,vo2,vo3}. They found a general correspondence between the emergent vielbein $e^i_\mu$ and the effective Hamiltonian near a Weyl node ${\bf K}=(K_x,K_y,K_z)$, i.e.,
{
\begin{equation}
H(\vec{q})=e_\mu^i q_i \sigma^\mu, \label{h1}
\end{equation}}
where $\mu=0,i$ with $i=x,y,z$. $q_i=k_i-K_i$ and  $\sigma^\mu=(I,\sigma^i)$ are the identity and Pauli matrices. Here the Einstein's summation convention for repeated indices is used. The 'speed of light'  $v_F$ (the Fermi velocity) , the electron {effective} mass $m_b$, and $\hbar$ are set to be {one} unless they are explicitly restored. The spectrum is given by
{
\begin{equation}
E(Q_i)_\pm=e_0^j [e^{-1}]_j^i Q_i\pm |\bf{Q}|, \label{spec}
\end{equation} }
with $Q_i\equiv e_i^j q_j$. The vielbein components $e^i_\mu$ together with $e^0_0=-1$ and $e^0_i=0$ define an emergent acoustic metric $g^{\mu\nu}=\eta^{\alpha\beta}e^\mu_\alpha e^\nu_\beta$ with the signature $\eta^{\mu\nu}={\rm diag}(-1,1,1,1)$ \cite{vo3}. {With this vielbein choice, the conical spectrum (\ref{spec}) can be written as
	\begin{equation}
	g^{\mu\nu}q_{\mu}q_{\nu}=0,
	\end{equation}
	where $q_0=E$ and the dual metric $g_{\mu\nu}$ defines the light cone with $g_{\mu\nu}x^\mu x^\nu=0$. Therefore,} the {ergosurface(ES)} of this emergent geometry is determined by {$g_{00}=-(1-\bf v^2)=0$ \cite{ergo}, i.e.,} $|{\bf v}|=1$ where $v^i=e_0^j [e^{-1}]_j^i$. A Weyl node is called type-I (type-II) if $|{\bf v}|<1$ ($|{\bf v}|>1$). Notice that the type of each one in a pairwise nodes can be arbitrary \cite{NN,hybrid}. {From this effective geometry point of view, quasi particles associated with the type-I (type-II) Weyl node effectively live outside (inside) the {ES}.}

\begin{figure}
	\includegraphics[width=0.4\textwidth]{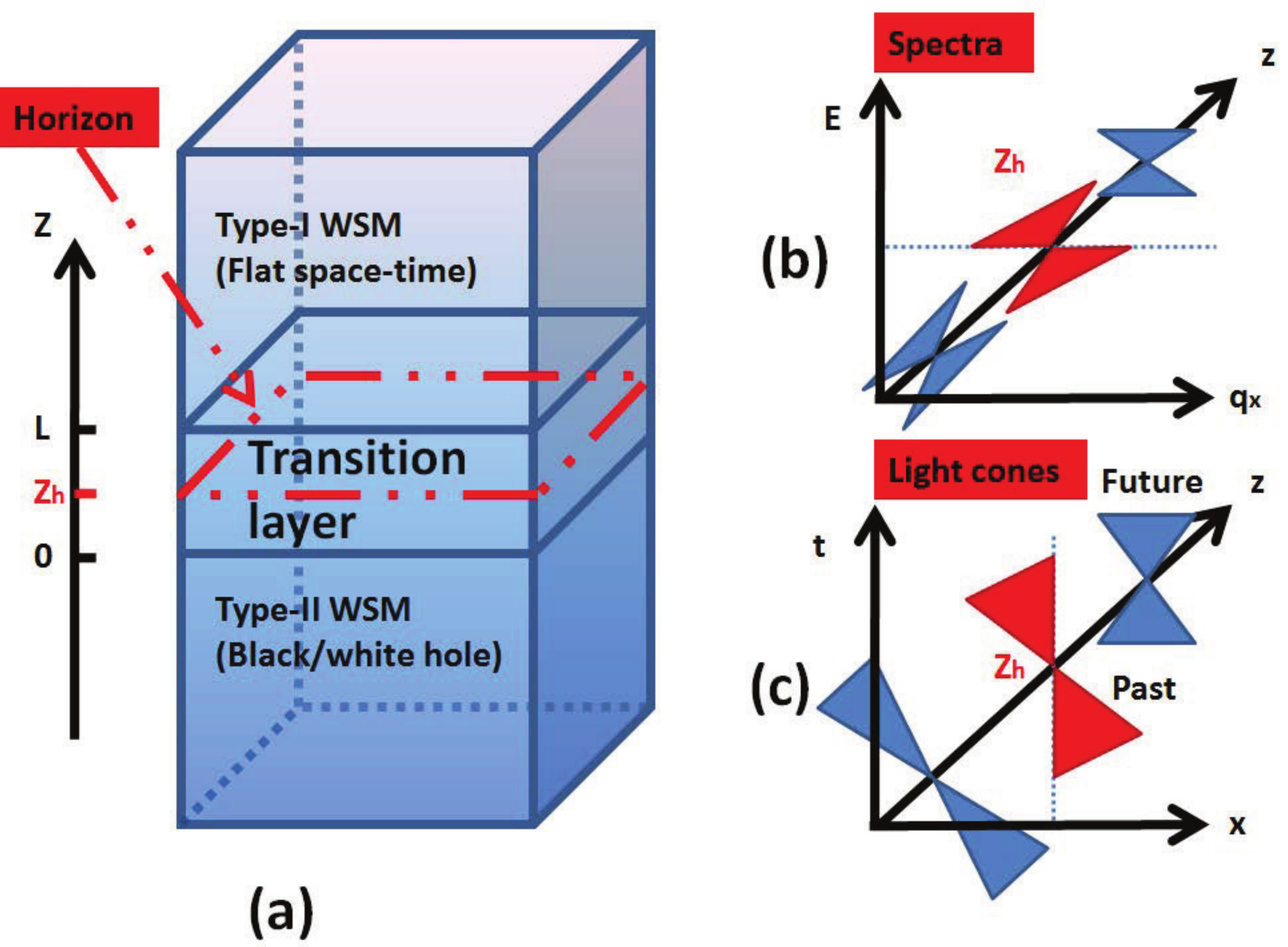}
	\caption{(color online)
	(a) The TL between Type-I and Type-II WSMs.  The acoustic event horizon is located at $z=z_h$ with $v_z(z_h)=1$.
	(b) The spectra outside, at, and inside the horizon, corresponding to type-I, critical and type-II WSM's. 
	(c) The light cones outside, at, and inside the horizon. 
	In (b) and (c), we take $\alpha_z(z)=0$ and project to the $x$-direction for simplicity. 
\label{fig1}	}
\end{figure}

Volovik uses the spherically symmetric metric to study inhomogeneous WSM \cite{vo1} {which} is difficult to be realized in reality. Both type-I and type-II WSMs have been found \cite{wan,weng,huang,xu,lv,yang,solu}. In order to simulate an acoustic event horizon {(which coincides with the ES when the normal velocity $v_\perp$ dominates \cite{ergo})} in a realizable geometry and study the chaos phenomena, we consider a type-I/type-II WSM transition layer (TL) where the type-I and type-II Weyl nodes may {mismatch} (See Fig. \ref{fig1}(a)). {In the following, we'll assume $v_{\perp}$ dominates whenever the horizon is mentioned.} Within the TL, a black/white hole planar horizon emerges effectively at a plane with $|{\bf v}_z|=1$ in the TL. A fermionic shock wave and the Hawking evaporation are possible to be detected because the Hawking temperature may be as high as several tens of Kelvin.    

In the effective Hamiltonian approach, we further find that when the Weyl nodes of the type-I and type-II do not completely match, the TL states inside and outside the {ES} may be totally different: On the type-I WSM side, the  TL state is described by the Landau levels of the WSM in an effective magnetic field.  On the type-II side, after a characteristic plane, the TL states are governed by an effective inverted oscillator potential \cite{bar,yuce,morita}, and thus are chaotic due to the non-integrability caused by coupling to the environment potential \cite{hash}. 

To further study the quantum chaos of the quasiparticles in the TL, we study a lattice model which reduces to the effective chaotic model in the TL in the long wave length limit.  We calculate out of time ordered  correlators (OTOCs) of the quantum states which characterize the quantum chaotic behavior \cite{Larkin,mald2,kit1,kit2,mald1,sac,she1,she2}. For a proper hopping strength in the TL with a sufficient layer thickness, we find that 
in a lower temperature, the OTOC does not show a fast scrambling over a short time while it does when the temperature is higher and  a quantum Lyapunov exponent is obtained. The OTOCs  saturate after the Ehrenfest time. These two properties of OTOC indicate the existence of quantum chaos inside the {ES} \cite{hashi}.

\section{ Effective model for TL}
 We consider that type-I and type-II WSMs couple as shown in Fig.\ref{fig1}(a). The TL is located between $z=0$ and $z=L$. Assuming ${\bf K}^{I}$ (${\bf K}^{II}$) is a Weyl node of the type-I (type-II) WSM in the regime $z>L$ ($z<0$), and $|{\bf K}^I-{\bf K}^{II}|\ll K_0$ where $K_0$ is the distance between a pair of Weyl nodes in the bulk. The effective Hamiltonian in the bulk of the type-I/II WSM is given by  
\begin{eqnarray}
H_{I/II}&=&(k_i-K_i^{I/II})\sigma^i-\alpha_{x,I/II} (k_x-K_x^{I/II})\nonumber\\
&-&\alpha_{z,I/II}((k_z-K_z^{I/II}),
\end{eqnarray}
where  $\boldsymbol{\alpha}_{I/II}=(\alpha_{x,I/II},0,\alpha_{z,I/II}) $ are parameters that tilt the Dirac cones. For the type-I WSM,  $|\boldsymbol\alpha_I|=\sqrt{\alpha^2_{x I}+\alpha^2_{z I}}<1$ and $|\boldsymbol\alpha_{II}|=\sqrt{\alpha^2_{x II}+\alpha^2_{z II}}>1$ for the type-II WSM. We further assume $|\Delta \boldsymbol\alpha|=|\boldsymbol\alpha_{II}-\boldsymbol\alpha_I|\ll1$. {The small difference in the Weyl nodes can be viewed as a soft edge between the WSMs. The soft edge is achieved by a small linear deformation between the lattice constants, which corresponds to a linear interpolation between the Weyl nodes in the lowest order. Therefore the} TL effective Hamiltonian can be approximated by a linear interpolation between $H_I$ and $H_{II}$ \cite{tch} 
 \begin{eqnarray}
H(z) =H_{II}+(H_I-H_{II})\frac{z}{L}, \label{heff}
\end{eqnarray}
where $0\leq z\leq L$. 
 Defining ${\bf q}={\bf k}-{\bf K}^{II}$ and $B_i=\frac{K_i^I-K_{i}^{II}}L=\frac{\Delta K_i}L$, 
	\begin{eqnarray}
	H(z)&=&(q_x+B_xz)\sigma^x+(q_y+B_yz)\sigma^y+(-i\partial_z+B_zz)\sigma^z\nonumber \\
	&&-\alpha_x(z)(q_x+B_xz)-\alpha_z(z)(-i\partial_z+B_zz)+... \label{h2}
	\end{eqnarray}
where  ``...'' are high order terms including $O(\Delta \boldsymbol K \cdot \Delta\boldsymbol\alpha)$ etc. $q_z$ is replaced by $-i\partial_z$ and $B_x(B_y)$ can be thought as an effective magnetic field in the $y(x)$-direction.  One can perform a gauge transformation to change $-i\partial_z+B_zz$ to $-i\partial_z$ if the wave function $\psi$ changes to $e^{-i\frac{B_zz^2}2}\psi$ accordingly. $\boldsymbol\alpha(z)$ is a linear interpolation like $H(z)$, i.e.,
$\boldsymbol\alpha(z)= \boldsymbol\alpha_{II}-\Delta\boldsymbol\alpha\frac{z}L.$ 
Since $|\Delta\boldsymbol\alpha|\ll1$, $\boldsymbol\alpha(z)$ is a slow varying vector function of $z$. $|\boldsymbol\alpha(z_c)|=1$ determines the critical surface with $z=z_c$ which separates the type-I and type-II WSM, where
\begin{eqnarray}
\frac{z_c}L=\frac{\Delta\boldsymbol{\alpha}\cdot \boldsymbol{\alpha}_{II}-\sqrt{(\Delta\boldsymbol{\alpha}\cdot \boldsymbol{\alpha}_{II})^2-(\Delta\boldsymbol{\alpha})^2(\boldsymbol\alpha^2_{II}-1)}}{(\Delta\boldsymbol\alpha)^2}\label{hzn}
\end{eqnarray}

\section{ Emergent geometry and horizon }
To see the emergent geometry, we first consider $B_i=\frac{\Delta K_i}L=0$ for simplicity. {Comparing the TL Hamiltonian (\ref{h2}) with the effective Hamiltonian (\ref{h1}), we can identify the  vielbein:   
 $e^x_0=-\alpha_x(z), e^y_0=-\alpha_y(z)=0, e^z_0=-\alpha_z(z)$ and $e^i_j=\delta^i_j$. Therefore the effective geometry in the TL is described by the line element of the induced metric \cite{vo3}
\begin{eqnarray}
ds^2&=&-(1-|{\bf v}(z)|^2)dt^2-2{v^i}(z) d{r_i}dt+(d{\bf r})^2, \nonumber \\
&=& -d^2t+(dr_i-v_idt)^2, \label{line}
\end{eqnarray}
with $v^i=e^i_0=-\alpha_i(z)$ and $|{\bf v}|=|\boldsymbol\alpha(z)|=\sqrt{\alpha^2_x+\alpha^2_z}$. The ES  which is denoted as $z=z_c$ is defined by $g_{00}=-(1-{\bf v}(z)^2)=0$ and the acoustic event horizon is defined by $v_z(z_h)=1$ \cite{ergo,matt}. In the limit $\frac{v_x^2}{v_z^2}\rightarrow 0$, the ES coincides with the event horizon which we denote as $z=z_h$. 
For instance, we sketch the spectra and light cones with $\alpha_z(z)=1$ 
in Fig. \ref{fig1} (b) and (c).  In that emergent metric, the Ricci scalar $R$ at the acoustic event horizon is given by
	\begin{widetext}
	\begin{equation}
	R_{z_h}=\frac{\Delta\alpha_x^2(-2(1-\alpha_{IIz})^2\Delta\alpha_x^2-4\alpha_{IIx}(1-\alpha_{IIz})\Delta\alpha_1\Delta\alpha_3+(1-2\alpha_{IIx}^2)\Delta\alpha_z^2)}{2\Delta\alpha_z^2},
	\end{equation}
	\end{widetext}
	which reduces to zero when $\Delta\alpha_x\rightarrow 0$. Furthermore, the surface gravity $\eta_H$ can be identified by introducing  $L^\mu=(1,v_x,0,0)$ which is the null vector field on the acoustic event horizon with the integral curves generating the horizon  \cite{matt},
\begin{equation}
	L^\nu\partial_\nu L^\mu|_{z=z_h}=\frac{\partial(1-v_z^2)}{2\partial z}L^\mu=\eta_HL^\mu,
\end{equation} 
namely, 
\begin{equation}
	\eta_H=\frac{\partial(1-v_z^2)}{2\partial z}|_{z=z_h}=\Delta\boldsymbol\alpha\cdot \boldsymbol\alpha(z_h)\frac{1}{L},
\end{equation}
The Hawking temperature of this emergent black hole horizon is defined by \cite{matt}
\begin{equation}
T_H=\frac{\hbar \eta_H }{2\pi k_B}=\frac{\Delta\boldsymbol\alpha\cdot \boldsymbol\alpha(z_h)\hbar v_F}{2L\pi k_B}, \label{HR}
\end{equation} 
where we have explicitly restored the Fermi velocity whose typical value for WSM is $v_F\sim 10^6$ m/s. If the thickness of the TL  $L\lesssim10$nm, $T_H\gtrsim 10^2\Delta\boldsymbol\alpha\cdot \boldsymbol\alpha(z_h)$ K. If $\Delta\boldsymbol\alpha\cdot \boldsymbol\alpha(z_h)\sim 0.1$, the 
 Hawking temperature may arrive at several tens of Kelvin which are experimentally reachable. 
 
 In these discussions, we have assumed the Fermi velocity $v_F=1$ is a constant in the low energy effective theory. If the interactions and quantum fluctuations are considered, $v_F$ may be renormalized or even spatial dependent. 
 To exactly calculate the  Hawking temperature, one needs to solve the Weyl equation with the emergent metric (\ref{line}), and then a Hawking evaporation may be observed by a thermal spectrum of the TL fermion  when the layer is forming \cite{vo1,Liu2}. Similar to the transsonic wave in the bosonic fluid \cite{unruh1,vo4}, we shall have a fermion shock wave with a velocity ${\bf v}(z)$, which will be presented elsewhere. 
  
\section{ Chaotic transition layer } 
To show the {chaotic behavior in the effective theory, we consider the Dirac equation with mismatching Weyl nodes $0<\Delta K_i\ll K_0$, 
\begin{eqnarray}
H(z)\psi(z,t)=i\dot\psi(t), \label{D1}
\end{eqnarray}
where $H(z)$ is given by Eq. (\ref{h2}) and $\psi^T=(\chi,\zeta)$ is a two-component spinor. Without loss of generality, we take $q_y=0$. Because $\boldsymbol\alpha(z)$ slowly varies, we neglect $\partial_z \boldsymbol\alpha(z)$ term.} Then, the dominant part of $\chi$'s equation in the large $\tilde z$ limit reads 
\begin{eqnarray}
\ddot\chi -\partial^2_{\tilde z}\chi+[\kappa\tilde z^2 +B^2_y(q_x^2-\frac{2\sqrt{1-\alpha_z^2}}{B_x}q_x\tilde z)]\chi+...=0,~~\label{E2}
 \end{eqnarray}
 where $\tilde{z}=\frac{q_x+B_xz}{B_x\sqrt{1-\alpha_z^2}}$ and $\kappa=(1-\boldsymbol\alpha^2)B_x^2+(1-\alpha_z^2)B_y^2$ (See {Appendix A} for ``...'' terms in more details). For $\kappa=\omega^2>0$ {with} a fixed $q_x$,  Eq. (\ref{E2}) coincides with the Landau levels of a Weyl fermion {outside the {ES}}. 

{The Eq. (\ref{E2}) will become an inverted oscillator when $\kappa<0$ \cite{bar,yuce,morita}. This can be satisfied for $|\boldsymbol\alpha|>1$ and $|\alpha_z|<1$ after a characterized plane $z_c$ inside the {ES (${\bf v}(z_c)^2=1$ if $B_y=0$)} \cite{note1}. An inverted oscillator is known to lead to the chaotic dynamics when it couples to the environment $V(x)$ \cite{zuret,miller,hash}. The detailed form of $V(x)$ is not important as long as the potential $V(x)$ is a confining one. We therefore} introduce an environment potential $V(x)$ and then Eq. (\ref{E2}) becomes
\begin{eqnarray}
&&\ddot\chi -\partial^2_{\tilde z}\chi-\lambda^2_L\tilde z^2\chi -B^2_y\partial_x^2 +\epsilon_3\tilde z^3\chi+\epsilon_4\tilde z^4\chi\nonumber\\
&&+2iB_y^2B_x^{-1}\sqrt{1-\alpha_z^2}\tilde z \partial_x\chi+V(x)\chi=0, \label{E3}
 \end{eqnarray}
 where $\lambda^2_L=(\boldsymbol\alpha^2_{II}-1)B_x^2+(\alpha^2_{z,II}-1)B_y^2>0$,
 $\epsilon_3\sim O(\Delta\boldsymbol{\alpha}\cdot\boldsymbol{\alpha}_{II})$, and 
 $\epsilon_4\sim O((\Delta\boldsymbol{\alpha})^2) >0$ (see {Appendix A}).
 If $V(x)=0$, Eq. (\ref{E3}) is integrable {with a positive Lyapunov exponent $\lambda_L$,} and basically describes an inverted oscillator. {To explicitly show the classical chaos, we study the steady state solution of (\ref{E3}) and treat $E^2={\cal E}$ as a classical energy. The Poincar{\'e} sections for the classical limit of (\ref{E3}) with $V(x)\sim x^2$ show that the Kolmogorov-Arnold-Moser tori
keep nice periodic orbits in lower ${\cal E}$ but break down as ${\cal E}$ raises, which indicates the existence of classical chaos. For details, see {Appendix A}.} 

At the end of this section, we'd like to point out that, in the Hawking radiation (\ref{HR}), the temperature is proportional to $|\alpha_z(z_h)|=v_F$, while the Lyapunov exponent remains positive as long as $|\alpha_x|>1$ even if $\alpha_z=0$.} The Hawking radiation is not directly related to the chaotic behavior in the TL, {In other words, the chaos survives even if the ES does not coincide with the event horizon.

\section{ Lattice model } {To further show the quantum chaos} in the TL, we study a lattice model on a cubic lattice with a lattice spacing $a=1$.  For simplicity, we do not consider the environment disorder {while the Lyapunov exponent remains positive}. The periodic boundary conditions are implicated while the thickness in the $z$-direction is finite. We label the sizes of type-I/II WSMs and the TL as $L_{I/II}$ and $L$, respectively. We consider minimal models describing the type-I/II WSMs on a periodic cubic lattice with Hamiltonian $H^l_{I/II}$ \cite{lah}
\begin{equation}
H^l_{I/II}=d _{I/II,\mu}\sigma^\mu, \label{eq16}
\end{equation} 
 where 
\begin{eqnarray}
 d_{I/II,x}&=&2\tilde t(\cos k_x-\cos K^{I/II}_x) +2\tilde t(1- \cos(k_y-K^{I/II}_y))\nonumber\\
 &+&\frac{\tilde t}2(\cos 3k_x-\cos 3K^{I/II}_x)+2\tilde t\gamma (1- \cos k_z),\nonumber \\
 d_{I/II,y}&=&2\tilde t \sin (k_y-K^{I/II}_y),\nonumber d_{I/II,z}=-2\tilde t \gamma \sin k_z,\nonumber\\
  d_{I,0}&=&0, d_{II,0}=-2\eta_{II}(\cos k_x-\cos K^{II}_x). 
\end{eqnarray}
Here  $\tilde t$ is the hopping strength. We introduce $\gamma$, which is a $z$-dependent parameter (see Eq. (\ref{hi})), to fine tune the hopping strength in the $z$-direction. {The high energy states which may be quantum chaotic and far away from the Weyl points can be moved near the Fermi level by tuning $\gamma$}. The Weyl nodes are located at ${\bf K}^{I/II}=(K^{I/II}_x,K^{I/II}_y, 0)$.  The lattice Hamiltonian we use here is given by  
\begin{equation}
H^l=\left\{
\begin{array}{ll}
 H^l_{II}, &  \gamma=\gamma_{II}, n_z\leq L_{II},  \\
 H^l_{II}+\xi (H^l_{I}-H^l_{II}), &  \gamma=\gamma_{TL}, n_z\in (L_{II},L+L_{II}],\\
H^l_{I}, &  \gamma=\gamma_{I}, n_z>L+L_{II}, \label{hi}
\end{array}
\right.
\end{equation} 
 where $n_z$ is the lattice site index in the $z$-direction and $\xi=\frac{n_z-L_{II}}L$.  In the long wave length limit,  the {effective} Hamiltonian in the TL reduces to the form of Eq. (\ref{heff}) up to a $z$-dependent function which leads to  a line of Weyl nodes (see {Appendix C}). 
  
Taking ${\bf K}^I=(0.54\pi,0.46\pi,0), {\bf K}^{II}=(0.5 \pi,0,0) $, $\eta_{II}=0.8\tilde t$ and considering a half filling lattice with $L=L_I=L_{II}=100$, we sketch the Fermi arcs and pockets projected to the $k_x$-$k_y$ plane {with} the colored regions in Fig. \ref{fig2}. There are the Fermi arcs on the surfaces of WSMs and the Fermi pockets in the type-II WSM as expected. In the TL, there are Fermi arcs that connect the type-I Weyl nodes' images {with the type-II ones} with the same chirality. The Weyl nodes' lines slightly deviate from these Fermi arcs  and the {ES} is at $\xi_h\approx 0.18$ (see {Appendix C}). According to the band structures of the lattice model (see {Appendix B}),  there is no  Fermi pocket in the TL for $\gamma_I=\gamma_{II}=\gamma_{TL}=1$ (Fig. \ref{fig2}(a)) while Fermi pockets emerge in the TL for $\gamma_I=\gamma_{II}=1$ and $\gamma_{TL}=0.7$ {by tuning $\gamma$} (Fig. \ref{fig2}(b)).

 \begin{figure}
 
 	\includegraphics[width=0.2\textwidth]{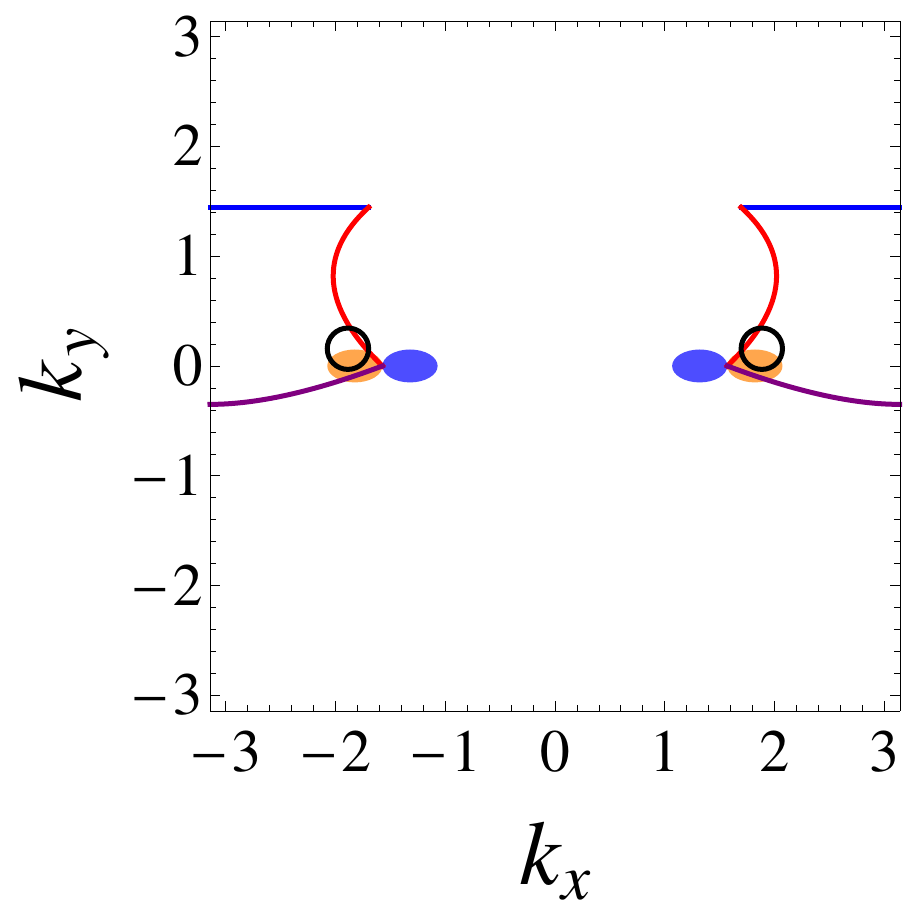}
	\includegraphics[width=0.2\textwidth]{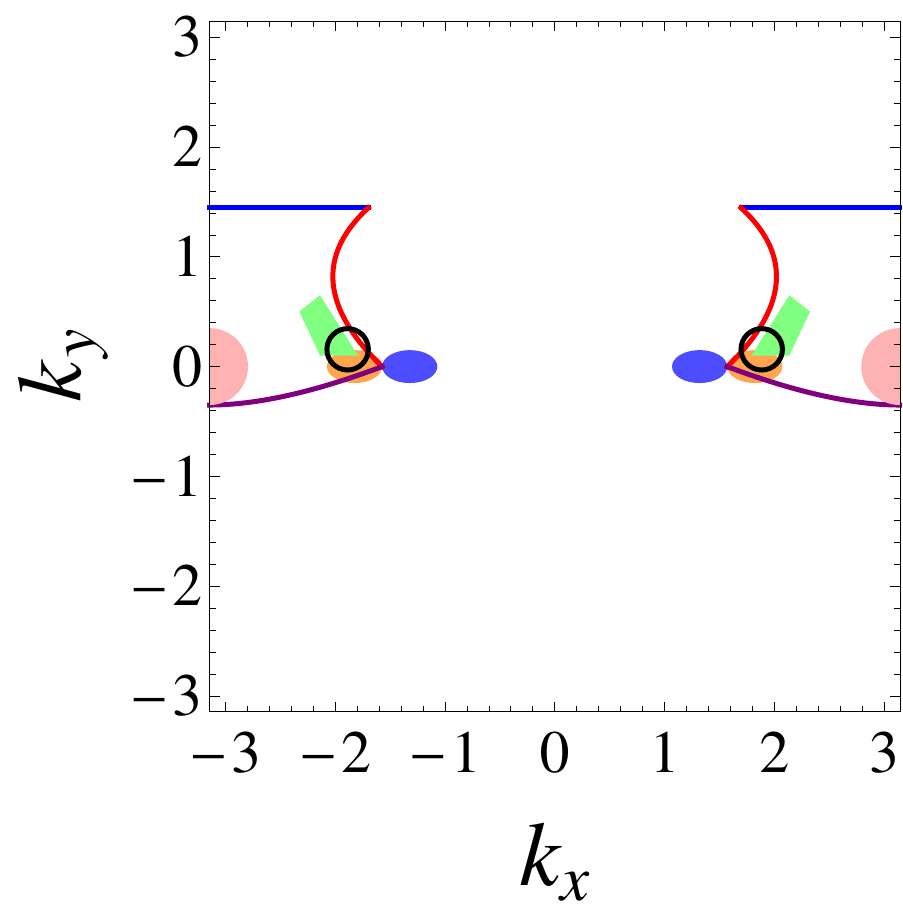}
	\centerline{~~~~~~(a)~~~~~~~~~~~~~~~~~~~~~~~~~~~~~~(b)}	
	 \caption{(color online) 
	The zero energy Fermi arcs and pockets in the $k_x$-$k_y$ plane.  (a) $\gamma_I=\gamma_{II}=\gamma_{TL}=1$. (b) $\gamma_I=\gamma_{II}=1$ and $\gamma_{TL}=0.7$. The red Fermi arcs lie in the TL and connect to the Weyl nodes's images of the WSMs.
	The orange and blue regions correspond to the quasiparticle and quasihole Fermi pockets in the type-II WSM. {Approximately, the states inside the black circle are described by the effective inverted oscillator and lie in the {ES}.} The additional areas in (b) are the pink quasiparticle pockets belonging to the interval inside the {ES} in the TL, and the green quasiparticle pockets are very close to the {ES} from the type-II WSM side. 
	  \label{fig2}}
\end{figure}

 \section{  OTOCs }  The sensitivity of the initial value in quantum chaos could be measured by
OTOCs \cite{mald2}.  We consider the simplest OTOC between the position and momentum operators in the $z$-direction which is measurable in the transport experiment \cite{Larkin}. In the semiclassical limit, this OTOC describes the dependence
on the initial condition of the particle motion.  On the lattice, these operators {corresponds to the site operator $n_z$ and the difference operator $\hat P_z$ which is defined by
	\begin{equation}
	 \hat P_z|\psi_{\bf k}(n_z)\rangle\equiv\frac{1}2(|\psi_{\bf k}(n_z-1)\rangle-|\psi_{\bf k} (n_z+1)\rangle),
	\end{equation}
	where  ${\bf k}=(k_x,k_y)$ and $|\psi_{\bf k}(n_z)\rangle$ is the state on site $n_z$.} The OTOCs are defined by 
 \begin{eqnarray}
 C_T(t)&=&\frac{1}Z{\rm Tr}(e^{-\beta H^l} [n_z(t),\hat P_z(0)]^2) \nonumber\\
 &=& \sum_{\psi}f_\psi\langle \psi|[n_z(t),\hat P_z(0)]^2|\psi\rangle,
 \end{eqnarray}
  where $n_z(t)$ and $\hat P_z(t)$ are the time-dependent operators  in Heisenberg's picture.  $f_\psi=\frac{1}{e^{\beta E_{\psi_{\bf k}}}+1}$ is the Fermi distribution function. More explicitly, the OTOCs are given by  
 \begin{eqnarray}
 C_T(t)&=&\sum_{\bf k} C_T(t,\psi_{\bf k}), \label{sover}
 \end{eqnarray}
 where the ${\bf k}$-dependent OTOC is given by 
 \begin{eqnarray}
  &&C_T(t,\psi_{\bf k}) =\sum _{\psi_1,\psi_2,\psi_3} f_\psi(1-f_{\psi_1}) \label{kfdC}  \\
  &&\times (1-f_{\psi_2})(1-f_{\psi_3}) \rho_{\psi\psi_1} \rho_{\psi_1\psi_2} \rho_{\psi_2\psi_3}\rho_{\psi_3\psi}  \nonumber \\
  &&\times[n_z(t)_{\psi\psi_1} P_z(0)_{\psi_1\psi_2}n_z(t)_{\psi_2\psi_3} P_z(0)_{\psi_3\psi}+\cdots ] ,\nonumber
    \end{eqnarray} 
 $O_{\psi\psi'}=\langle \psi |\hat O|\psi'\rangle$ for $\hat O=n_z(t)$ and $\hat P_z(0)$. $\rho_{\psi\psi'}$ is the probability that the state $|\psi'\rangle$ transits to the state $|\psi\rangle$ due to  a disturbance. For convenience, we take
 $\rho_{\psi\psi'}=e^{-\beta(E_\psi-E_{\psi'})}$ for $E_\psi>E_{\psi'}$ or 1 otherwise;  $O_{fi}=\langle f |\hat O|i\rangle$ for $\hat O=n_z(t)$ and $\hat P_z(0)$.  $"\cdots"$ is the other three terms when explicitly writing out $[n_z(t),\hat P_z(0)]^2$.    

 \begin{figure}
	\includegraphics[width=0.2\textwidth]{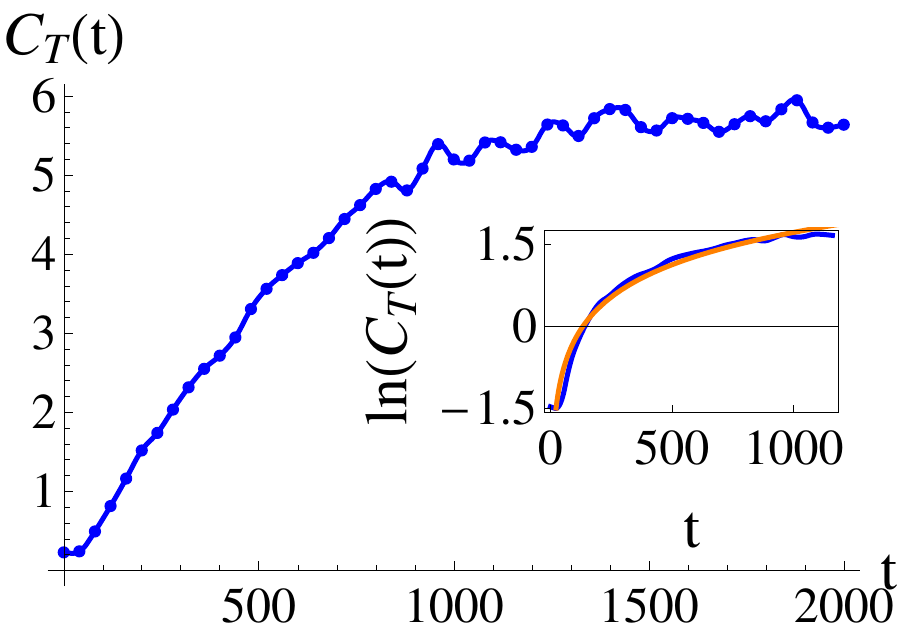}
	\includegraphics[width=0.2\textwidth]{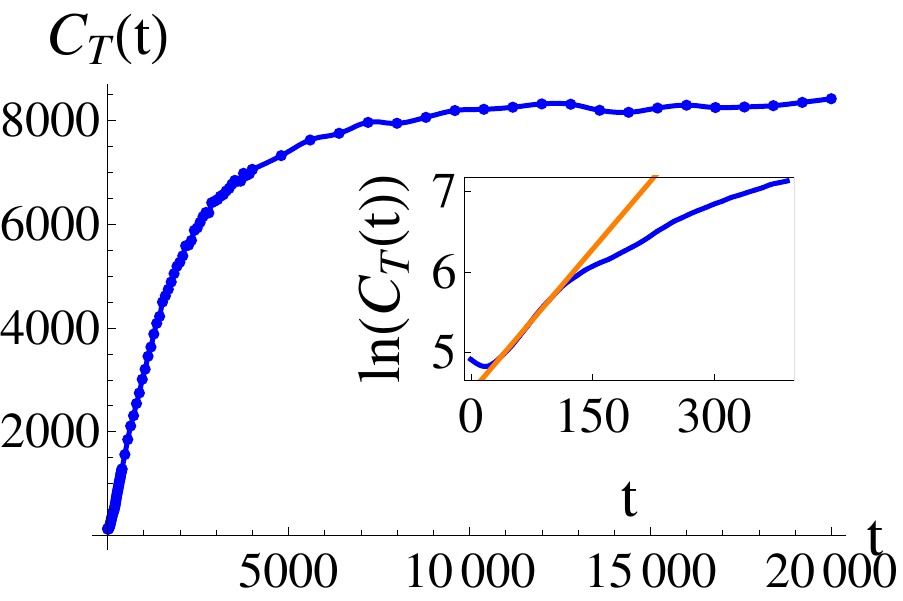}
	\centerline{(a)~~~~~~~~~~~~~~~~~~~~~~~~~~~~~(b)}
	\includegraphics[width=0.2\textwidth]{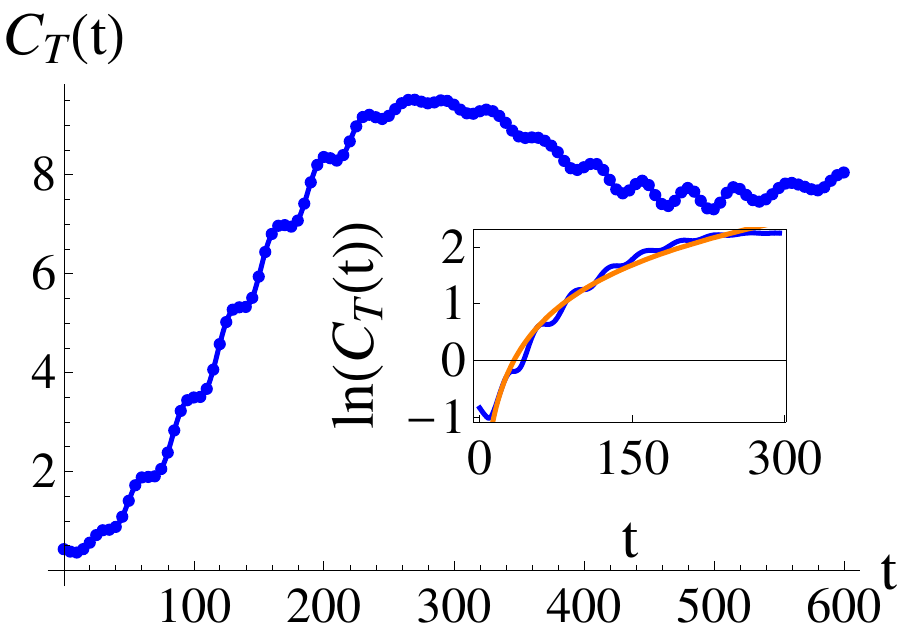}
	\includegraphics[width=0.2\textwidth]{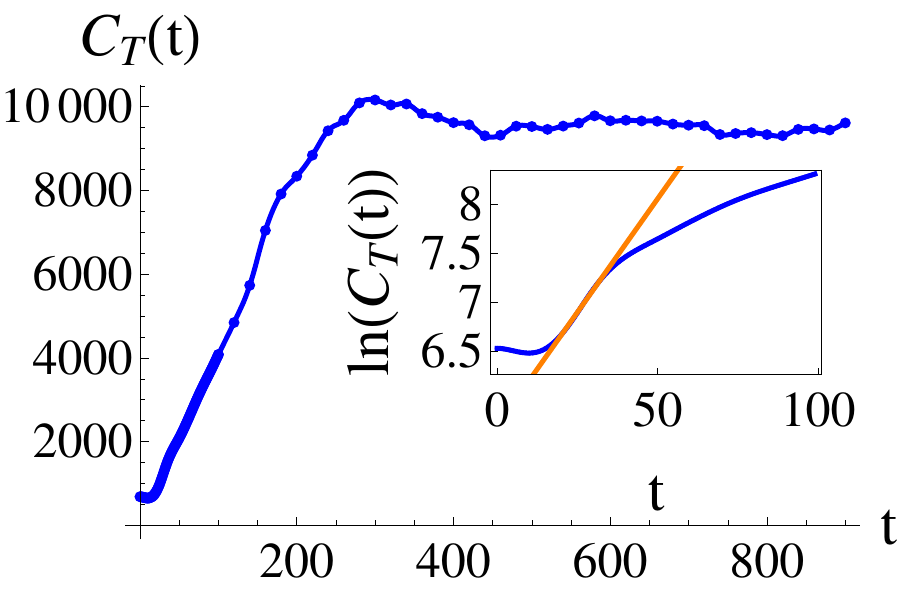}
	\centerline{(c)~~~~~~~~~~~~~~~~~~~~~~~~~~~~~(d)}	
			\caption{(color online) The OTOCs $C_{T}(t)$. (a) $\gamma_{TL}=1$ and $T=\tilde t/4$.   (b)  $\gamma_{TL}=1$ and $T=2\tilde t/3$. (c)  $\gamma_{TL}=0.7$ and $T=\tilde t/4$.   (d) $\gamma_{TL}=0.7$ and $T=2\tilde t/3$. 
				{The insets are a logarithmic fitting (red curve) of the $\ln C_T(t)$ in the short time for (a) and (c), while the ones in (b) and (d) includes the linear region (red line) of the $\ln C_T(t)$ with $\lambda_{QL}=0.006$ and $\lambda_{QL}=0.023$ respectively. }
					\label{fig3}	}
		\end{figure}

In numerical calculations, we take a lower temperature $T=1/\beta=\tilde t/4$ and a higher $T=2\tilde t/3$.   For the model with $\gamma_I=\gamma_{II}=\gamma_{TL}=1$, the OTOC with the lower $T$ has a small magnitude and does not have an exponential fast scrambling, as show in Fig. \ref{fig3}(a). In the inset in Fig. \ref{fig3}(a),{ the negative slope of the $\ln C_T(t)$ in the very short time} implies an exponential decay of the OTOC. {The reason lies in the binding potential of the quasiparticles on the lattice site at the beginning. After that time period, $\ln C_T(t)$ behaves logarithmic.} As the temperature raises,  the fast scrambling of the OTOC appears and its magnitude  becomes three order larger than that in the lower temperature as shown in Fig. \ref{fig3}(b).  In the inset of Fig. \ref{fig3}(b), {the OTOC still decays exponentially in a very short time.} After this very short time period, a time-dependent positive quantum Lyapunov exponent  $\lambda_{QL}(t)$ {can be read out} according to an exponential fitting $ C_T(t)\sim e^{2\lambda_{QL}(t) t}$.  There is a time period $t_Q\approx 100$ in which $\lambda_{QL}=0.006$ is constant \cite{note}.  This implies the fast scrambling of the OTOC. {The emergence of the quantum chaotic behavior} in the higher temperature is consistent with the classical chaos in the effective model because the chaos appears at higher energy. After $t_Q$, due to the quantum fluctuation, the fast scrambling begins to be suppressed and $\ln C_T(t)$ deviates from the linear fitting. After a long time $t_E\gtrsim 5000$, the OTOC is saturated instead of growing everlastingly in the classical chaos.  This time $t_E$ corresponds to the Ehrenfest time. This shows a possible distinction between the quantum chaos and the classical one \cite{hashi}. 

 For $\gamma_I=\gamma_{II}=1$ and $\gamma_{TL}=0.7$, as shown in Fig. \ref{fig3} (c) and (d),  the fast scrambling in the OTOC {is absent} for the lower temperature while it {exists} in the higher temperature, similar to that for $\gamma_{TL}=1$. However, the time $t_Q\approx 40$ and the saturated time $t_E\approx300$ 
are  shorter than those for $\gamma_{TL}=1$. This is because the green pockets emerge in Fig. \ref{fig2}(b), which increases the density of states and then enhances the quantum fluctuation.   Meanwhile, the higher density of states also raises the quantum Lyapunov exponent in linear region about 4 times, $\lambda_{QL}=0.023$. This means that the scrambling is much faster than that for $\gamma_{ TL}=1$. {In both cases of $\gamma_{TL}=1$ and $\gamma_{TL}=0.7$, the OTOCs are dominated by the states inside the black circle which corresponds to the inverted oscillator in the effective theory. Besides the black circle area, there are other states contribute to the positive Lyapunov exponent, such as the pink area in Fig. \ref{fig2}(b) which emerge by tuning $\gamma_{TL}$. These states are not predicted by the effective Hamiltonian (\ref{heff}).}
 
We have yet to consider the $x$-$y$ plane environment potential  in the lattice model.  The integrability of the model is lost if it is turned on and then the fast scrambling of the OTOC and the Ehrenfest time identify the quantum chaos.  

\section{ Conclusions }  We pointed out that there are emergent {ESs which coincides with acoustic event horizons under certain conditions} in the TL between type-I and type-II WSMs. {And the corresponding Hawking temperature may be measurable.}  When the Weyl nodes of type-I and type-II mismatch, the quasiparticles moving inside the {ES} may be quantum chaotic, {even if the event horizon and ES do not coincide.} We confirm this quantum chaotic behavior in a lattice model by calculating the OTOCs of the system. Two diagnostic quantities, the quantum Lyapunov exponent and the Ehrenfest time, which characterize the quantum chaos, were determined.

\acknowledgments

 This work is supported by NNSF of China with No. 11474061 (YGC,XL,YY), No. 11804223 (XL) and the ministry of science and technology of China with the grant No.2016YFA0301001 (FYL).

\clearpage

\begin{widetext}

\appendix

\section{Details on solving Eq. (\ref{D1}) in the main text and its classical Poincar{\'e} section}

The equations of motion (\ref{D1}) in the main text  for the wave function $\psi=(\chi,\lambda)^T$ read
\begin{eqnarray}
&&(q_x+B_xz)\lambda - i(q_y+B_yz) \lambda+(-i\partial_z+B_zz)\chi-\alpha_x(q_x+B_xz)\chi-\alpha_z(z)(-i\partial_z+B_zz)\chi=i\partial_t\chi,\label{a1}\\
&&(q_x+B_xz)\chi + i(q_y+B_yz) \chi-(-i\partial_z+B_zz)\lambda-\alpha_x(q_x+B_xz)\lambda-\alpha_z(z)(-i\partial_z+B_zz)\lambda=i\partial_t\lambda.\label{a2}
\end{eqnarray}
Without loss of generality, we take $q_y=0$. Notice that the $B_zz$ term is a pure gauge and can be gauged away by a gauge transformation $\psi\rightarrow \exp(-i\frac{B_zz^2}{2})\psi$. Solving Eq. (\ref{a1}), we have $\lambda=\lambda(\chi)$ and then substitute it  into Eq. (\ref{a2}),
\begin{eqnarray}
&&(((1-\alpha_x^2)(q_x+B_xz)^2+B_y^2z^2)-\partial_z^2+2i\alpha_x(q_x+B_xz)(\alpha_z \partial_z-\partial_t)-2\alpha_z \partial_z\partial_t+\alpha_z^2\partial_z^2+\partial_t^2)\chi\nonumber\\
&&-\frac{1}{q_x+(B_x-iB_y)z}(\alpha_x(1+\alpha_z)B_yq_x+(\alpha_z^2B_x-B_x+iB_y-i\alpha_z^2B_y)\partial_z+((1-\alpha_z)B_x+i(1+\alpha_z)B_y\partial_t))\chi=0. \label{a3}
\end{eqnarray}
When $q_x+B_xz>>1$, the second line of Eq. (\ref{a3}) can be neglected. Making a transformation $\chi\rightarrow \exp(i\frac{\alpha_x\alpha_z (q_x+B_xz)^2}{2B_x(1-\alpha_z^2)})\chi$, Eq. (\ref{a3}) becomes,
\begin{eqnarray}
&&(i\alpha_x\alpha_z(1-\alpha_z^2)B_x+\alpha_x^2(q_x+B_xz)^2-(1-\alpha_z^2)((q_x+B_xz)^2+B_y^2z^2))\chi\nonumber\\
&&+(1-\alpha_z^2)^2\partial_z^2\chi+2i\alpha_x(q_x+B_xz)\partial_t\chi+2\alpha_z(1-\alpha_z^2)\partial_t\partial_z\chi-(1-\alpha_z^2)\partial_t^2\chi=0.\label{a4}
\end{eqnarray}
Defining $\tilde{z}=\frac{q_x+B_xz}{Bx\sqrt{1-\alpha_z^2}}$, Eq. (\ref{a4}) reads,
\begin{eqnarray}
&&\partial_t^2\chi-\partial_{\tilde{z}}^2\chi+((1-\alpha_x^2-\alpha_z^2)B_x^2\tilde{z}^2+B_y^2(\sqrt{1-\alpha_z^2}\tilde{z}-\frac{q_x}{B_x})^2)\chi\nonumber\\
&&-i\alpha_x\alpha_zB_x\chi-2\alpha_z\sqrt{1-\alpha_z^2}\partial_0\partial_{\tilde{z}}\chi-\frac{2i\alpha_xB_x\tilde{z}}{\sqrt{1-\alpha_z^2}}\partial_t\chi=0.\label{a5}
\end{eqnarray}
Eq. (\ref{a5}) describes a harmonic oscillator when $\omega^2=(1-\alpha_x^2-\alpha_z^2)B_x^2+(1-\alpha_z^2)B_y^2>0$ or an inverted one with positive Lyapunov exponent $\lambda_L$ when $-\lambda^2_L=(1-\alpha_x^2-\alpha_z^2)B_x^2+(1-\alpha_z^2)B_y^2<0$. The leading order terms are the first line of Eq. (\ref{a5}), and the last three terms only {shift the oscillating center and the zero point energy.} Since we have treated the slow-varying functions ${\boldsymbol \alpha}={\boldsymbol \alpha_{II}-\Delta {\boldsymbol \alpha}\frac{z}{L}}$ as constants in the derivation of Eq. (\ref{a5}), the Lyapunov exponent $\Lambda_L$ can be expressed as
\begin{eqnarray}
\lambda_L^2(\tilde{z})&=&(\boldsymbol \alpha_{II}^2-2\boldsymbol \alpha_{II}\cdot\Delta {\boldsymbol \alpha}\frac{z}{L}+(\Delta\boldsymbol \alpha\frac{z}{L})^2-1)B_x^2+(\alpha_{zII}^2-2\alpha_{zII}\Delta\alpha_z\frac{z}{L}+(\Delta\alpha_z\frac{z}{L})^2-1)B_y^2\nonumber\\
&=&[(\boldsymbol \alpha_{II}^2-1)+(\frac{2\boldsymbol{\alpha}_{II}\cdot\Delta\boldsymbol{\alpha}q_x}{B_xL}+\frac{\Delta\boldsymbol{\alpha}^2q_x^2}{B_x^2L^2})-\frac{2\sqrt{1-\alpha_{zII}^2}(B_x\Delta\boldsymbol{\alpha}\cdot\boldsymbol{\alpha}_{II}L+\Delta\boldsymbol{\alpha}^2q_x)\tilde{z}}{B_xL^2}+\frac{(\alpha_{zII}^2-1)\Delta\boldsymbol{\alpha}^2\tilde{z}^2}{L^2}]B_x^2\nonumber\\
&&+[(\alpha_{zII}^2-1)+(\frac{2{\alpha}_{zII}\Delta{\alpha_z}q_x}{B_xL}+\frac{\Delta{\alpha_z}^2q_x^2}{B_x^2L^2})-\frac{2\sqrt{1-\alpha_{zII}^2}(B_x\Delta{\alpha_z}{\alpha}_{zII}L+\Delta{\alpha_z}^2q_x)\tilde{z}}{B_xL^2}+\frac{(\alpha_{zII}^2-1)\Delta{\alpha_z}^2\tilde{z}^2}{L^2}]B_y^2.\nonumber\\
\end{eqnarray}  
Therefore the chaos related terms in Eq. (\ref{a5}) reads
\begin{equation}
\partial_t^2\chi-\partial_{\tilde{z}}^2\chi-\lambda_{LII}^2\tilde{z}^2\chi-2B_y^2B_x^{-1}\sqrt{1-\alpha_z^2}q_x\tilde{z}\chi+B_y^2B_x^{-2}q_x^2\chi+\epsilon_3\tilde{z}^3\chi+\epsilon_4\tilde{z}^4\chi+...=0, \label{a7}
\end{equation}
where
\begin{eqnarray}
&&\epsilon_3=\frac{2\sqrt{1-\alpha_{zII}^2}((B_x\Delta\boldsymbol{\alpha}\cdot\boldsymbol{\alpha}_{II}L+\Delta\boldsymbol{\alpha}^2q_x)B_x^2+(B_x\Delta{\alpha_z}{\alpha}_{zII}L+\Delta{\alpha_z}^2q_x)B_y^2)}{B_xL^2},\\
&&\epsilon_4=\frac{(1-\alpha_{zII}^2)(\Delta\boldsymbol{\alpha}^2B_x^2+\Delta\alpha_z^2B_y^2)}{L^2}.	
\end{eqnarray}
The first three terms in Eq. (\ref{a7}) represent an inverted oscillator if $\Lambda_{LII}^2>0$. The $\tilde{z}^3$ and $\tilde{z}^4$ terms are the quantum perturbations, and the $q_x\tilde{z}$ term is the coupling between $\tilde{z}$ and $q_x$, the ``environment''. Therefor Eq. (\ref{a7}) can be viewed as an intrinsic quantum open system and chaos raises. 

To be more specific about the emergence of chaos, we consider the poincar{\'e} section of the classical limit of Eq. (\ref{a7}) up to the $\tilde{z}^4$ terms, i.e., the classical Hamiltonian reads,
\begin{equation}
H=q_{\tilde{z}}^2-\lambda_{LII}^2\tilde{z}^2+\epsilon_3\tilde{z}^3+\epsilon_4\tilde{z}^4-2B_y^2B_x^{-1}\sqrt{1-\alpha_z^2}q_x\tilde{z}+B_y^2B_x^{-2}q_x^2+\epsilon_xx^2, \label{a10}
\end{equation}
where $q_{\tilde{z}}$ is the momentum for $\tilde{z}$, and we also add a harmonic potential along the $x$-direction which simulates the effects of disorders. We draw its Poincar{\'e} section numerically in Fig. \ref{figPS} which clearly shows the existence of classical chaos. 
% Under these parameters, for $\alpha_x^2<2$ ($\Lambda_{LII}^2<0$), the Hamiltonian (\ref{a10}) is a harmonic oscillator, and its Poincar{\'e} section is a closed circle. And, for $\alpha_x^2>2$ ($\Lambda_{LII}^2>0$), the Hamiltonian (\ref{a10}) is a classical inverted oscillator, and its Poincar{\'e} section is an open curve, which indicates the existence of classical chaos.  

\begin{figure}
	\begin{minipage}{0.3\textwidth}
		\centerline{\includegraphics[width=1\textwidth]{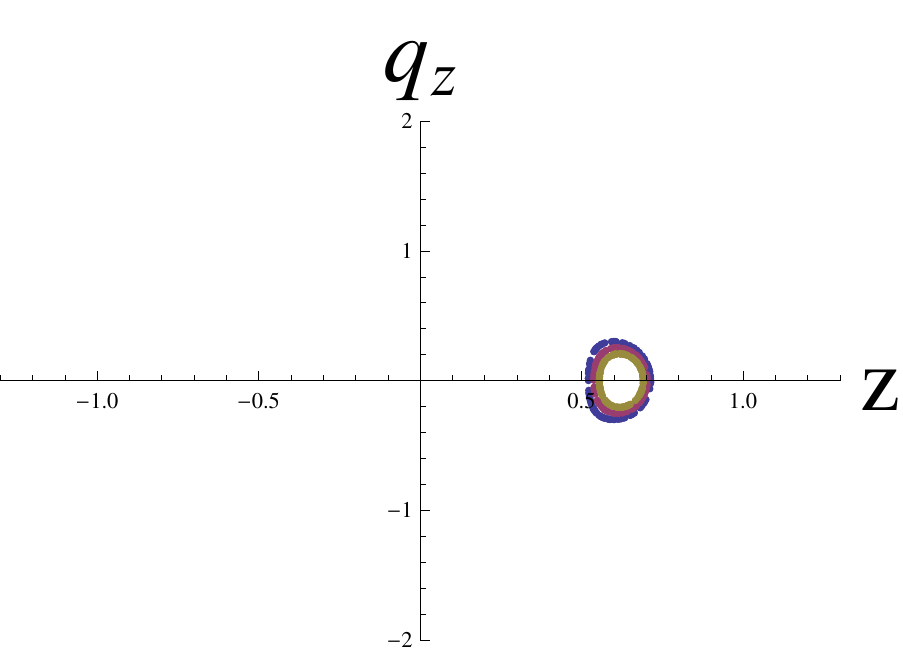}}%{poincaresection1.eps}}
		\centerline{(a)}
	\end{minipage}
	\begin{minipage}{0.3\textwidth}
		\centerline{\includegraphics[width=1\textwidth]  {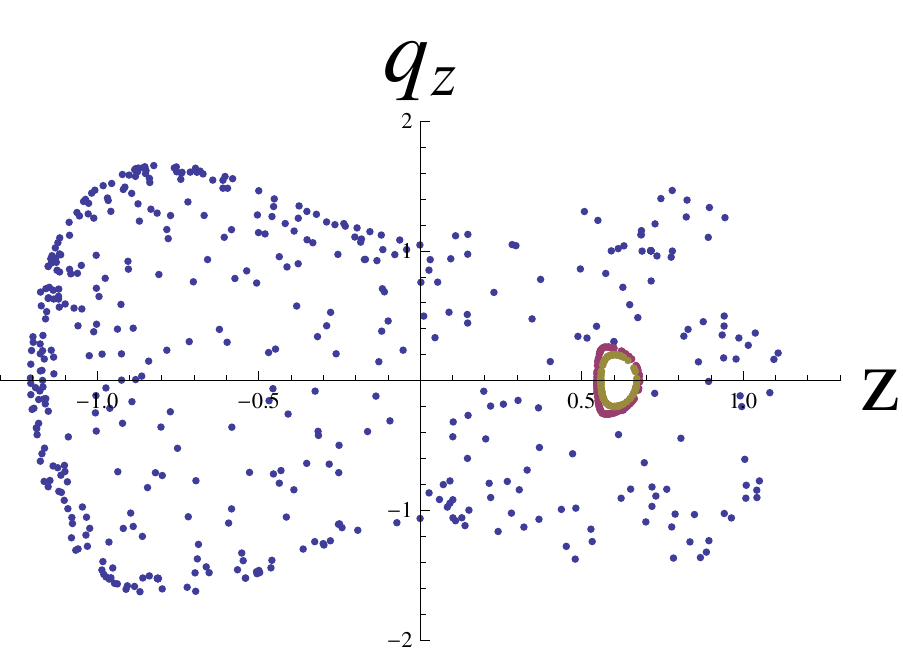}}% { poincaresection2.eps}}
		\centerline{(b)}
	\end{minipage}
	\begin{minipage}{0.3\textwidth}
		\centerline{\includegraphics[width=1\textwidth] {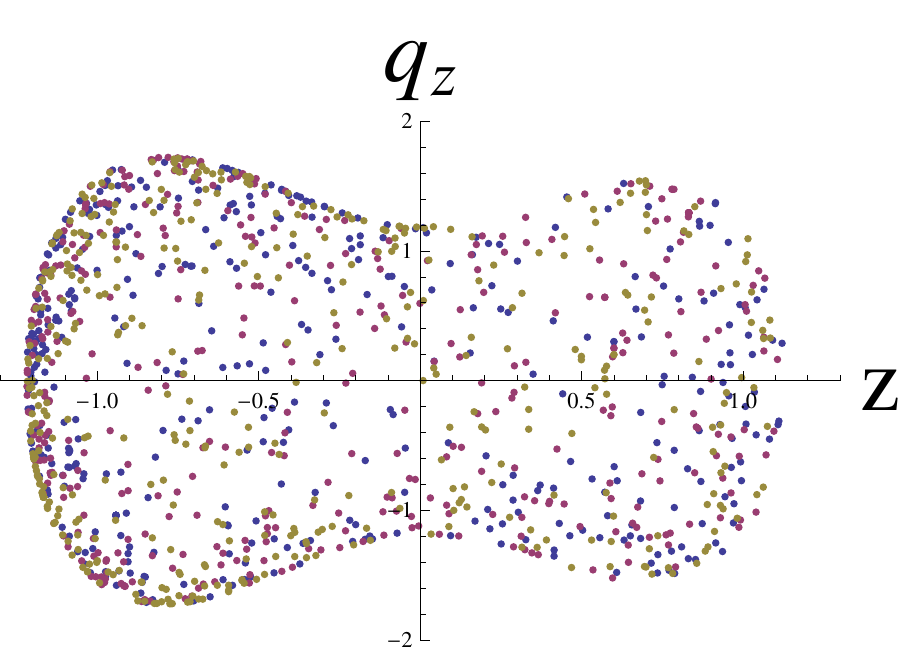}}% {poincaresection3.eps}}
		\centerline{(c)}
	\end{minipage}
	\caption{The Poincar{\'e} section for Hamiltonian (\ref{a10}). For simplicity, we choose $B_x=-1$, $B_y=1$, $\alpha_x=2$, $\alpha_z=0$, $\epsilon_3=0$, $\epsilon_4=4$, and $\epsilon_x=4$. The colors in the diagrams represent the KAM tori for different initial conditions: $q_z(0)=0.3,0.25,0.2$, $z(0)=0.6$, and $x(0)=0$. The energies are chosen: (a) $E=1$, (b) $E=1.2$, and (c) $E=1.4$. The breaking down of the KAM tori indicates the existence of classical chaos. 
		\label{figPS}	}
\end{figure}

\section{Band structure and Fermi energy}

We  give the band structures of the lattice model. Fig. \ref{fig5}(a) is for $\gamma_I=\gamma_{II}=\gamma_{TL}=1$ and Fig. \ref{fig5}(b) is for $\gamma_I=\gamma_{II}=\gamma_{TL}=1$. Corresponding zero energy Fermi arcs and pockets projected to the $k_x$-$k_y$ are shown in Fig. \ref{fig5}(c) and Fig. \ref{fig5}(d), respectively and they are {the original data for} sketching Fig. \ref{fig2}(a) and (b) in {the} main text.

\begin{figure}
	\begin{minipage}{0.3\textwidth}
		\centerline{\includegraphics[width=1\textwidth]{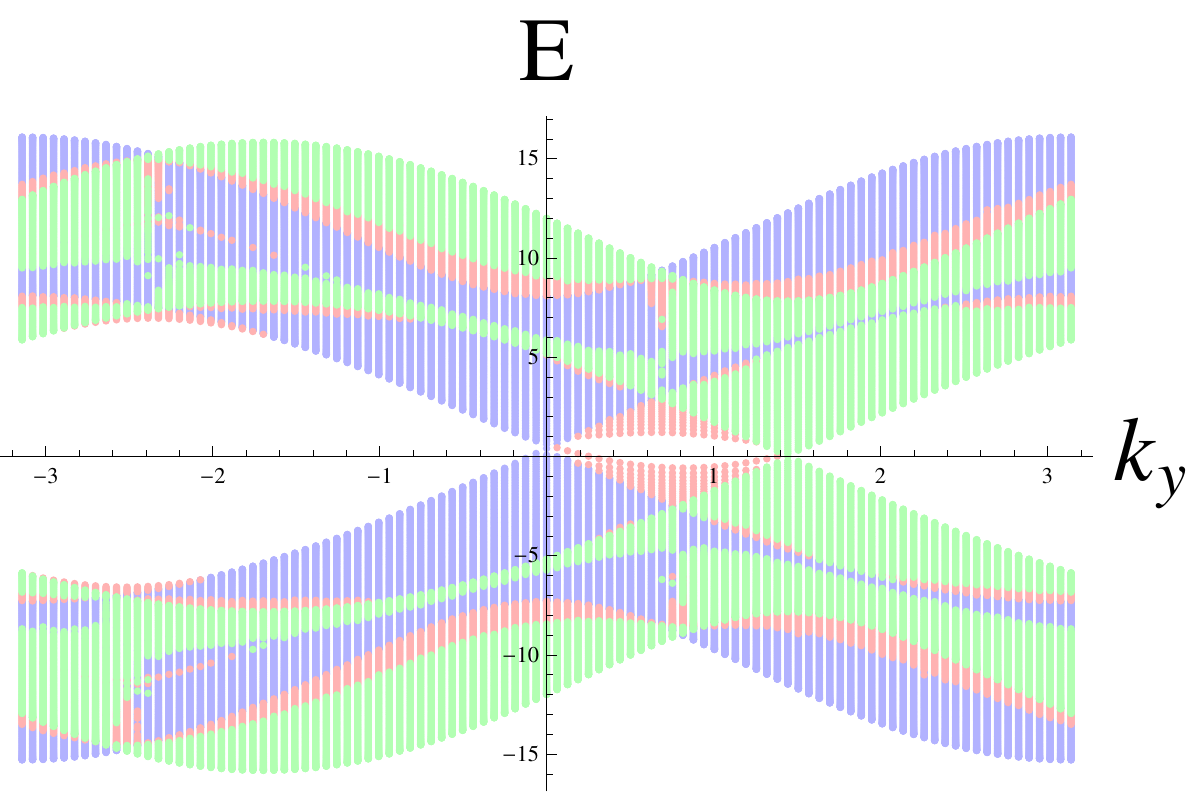}% {figsm1a1.eps}
			\includegraphics[width=1\textwidth]{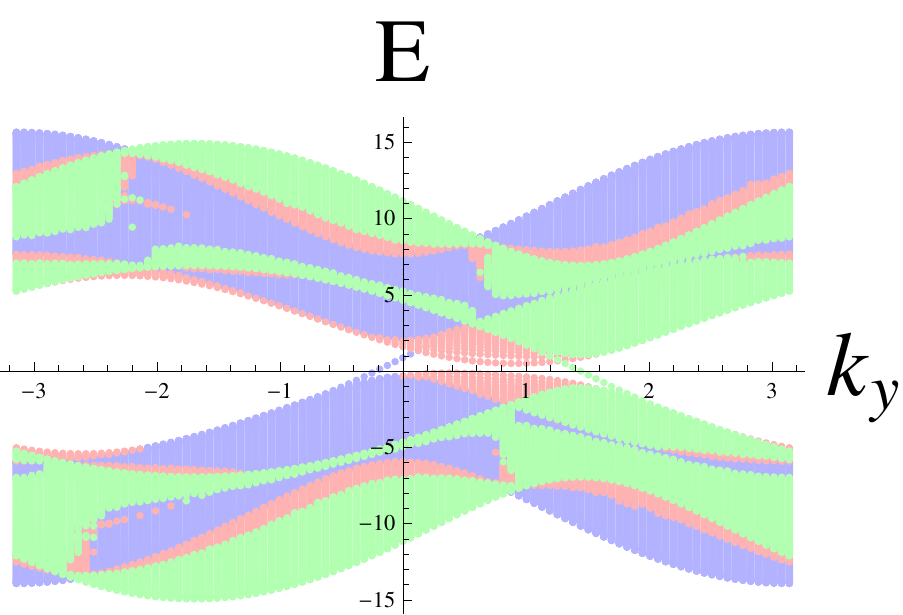}% {figsm1a2.eps}
			\includegraphics[width=1\textwidth] {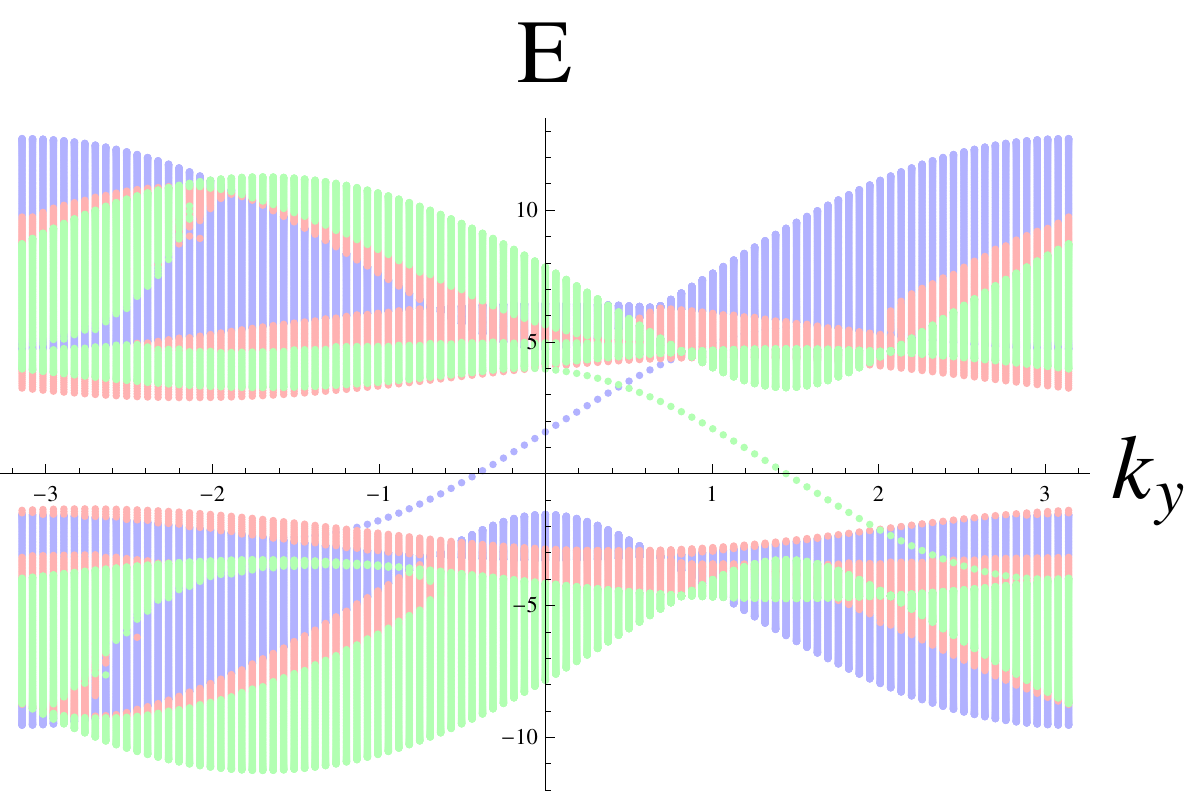}}%{figsm1a3.eps}}
		\centerline{(a)}
	\end{minipage}
	\\
	\begin{minipage}{0.3\textwidth}
		\centerline{\includegraphics[width=1\textwidth]{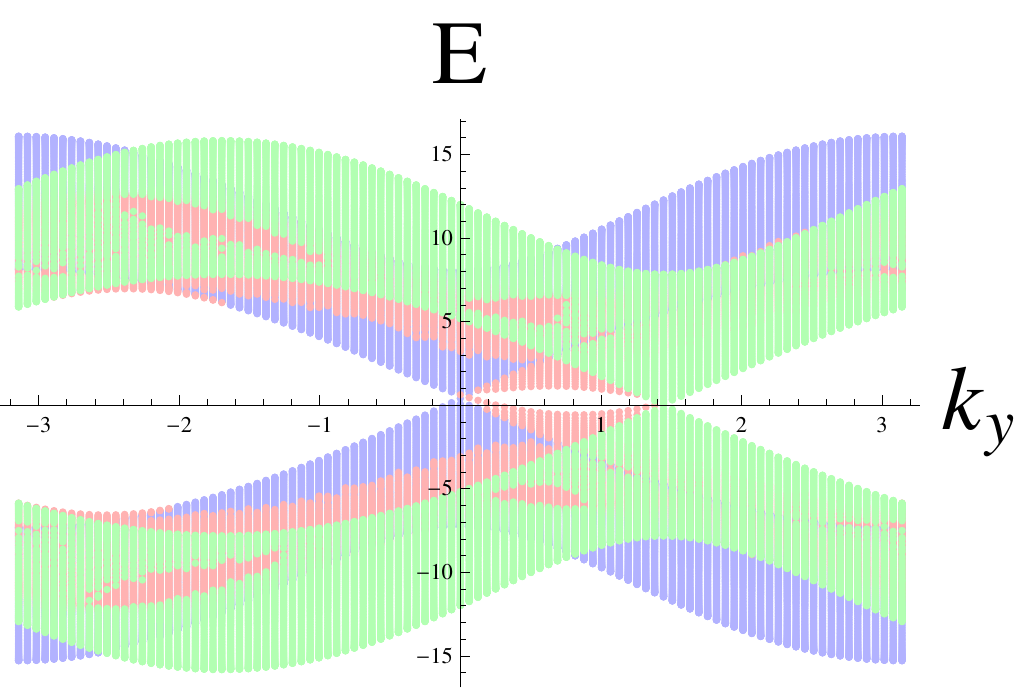}%{figsm1b1.eps}
			\includegraphics[width=1\textwidth] {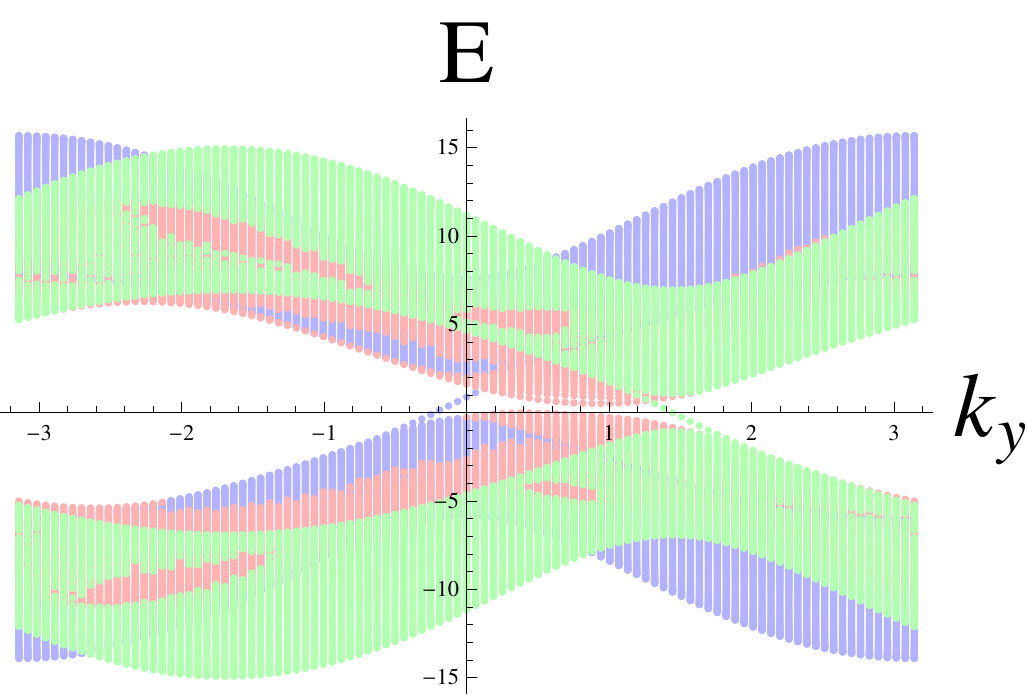}% {figsm1b2.eps}
			\includegraphics[width=1\textwidth] {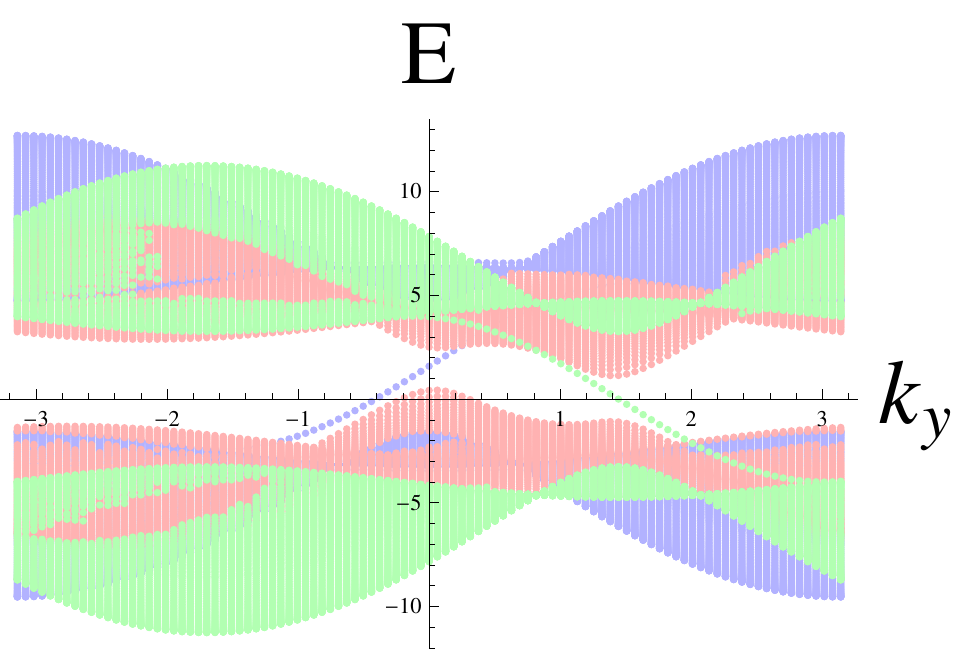}}%{figsm1b3.eps}}
		\centerline{(b)}
	\end{minipage}\\
	\begin{minipage}{0.4\textwidth}
		\centerline{\includegraphics[width=1\textwidth] {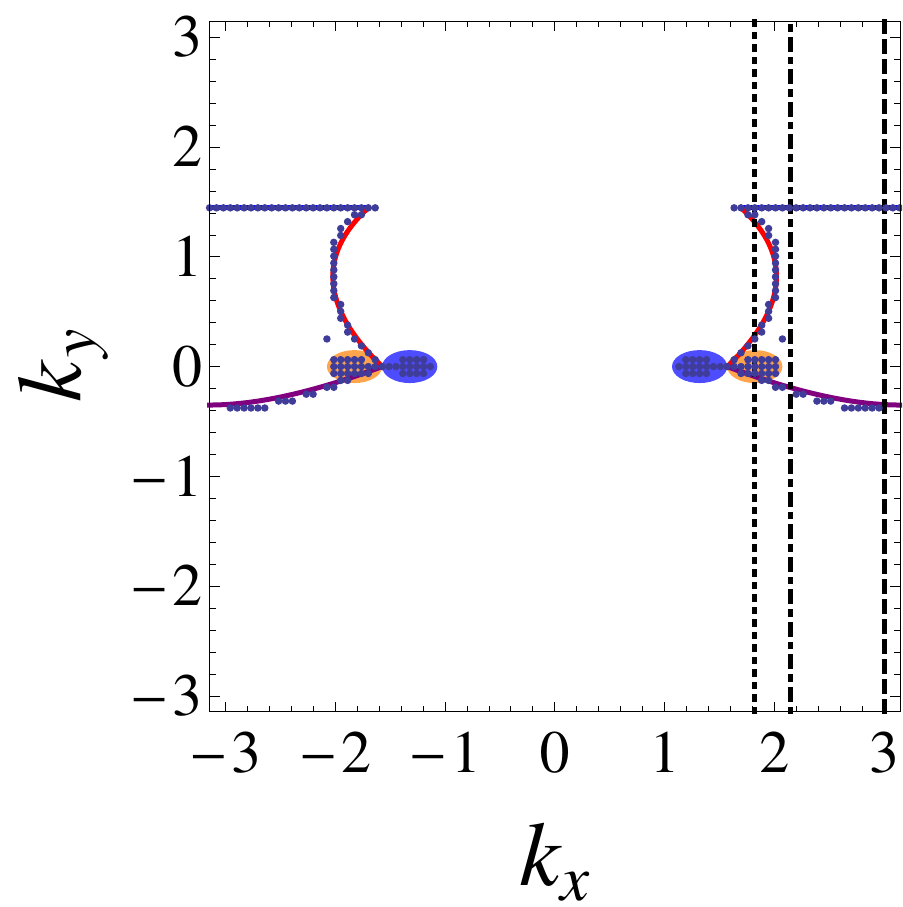}}%{figsm1c1.eps}}
		\centerline{(c)}
	\end{minipage}
	%	\hfill
	\begin{minipage}{0.4\textwidth}
		\centerline{\includegraphics[width=1\textwidth] {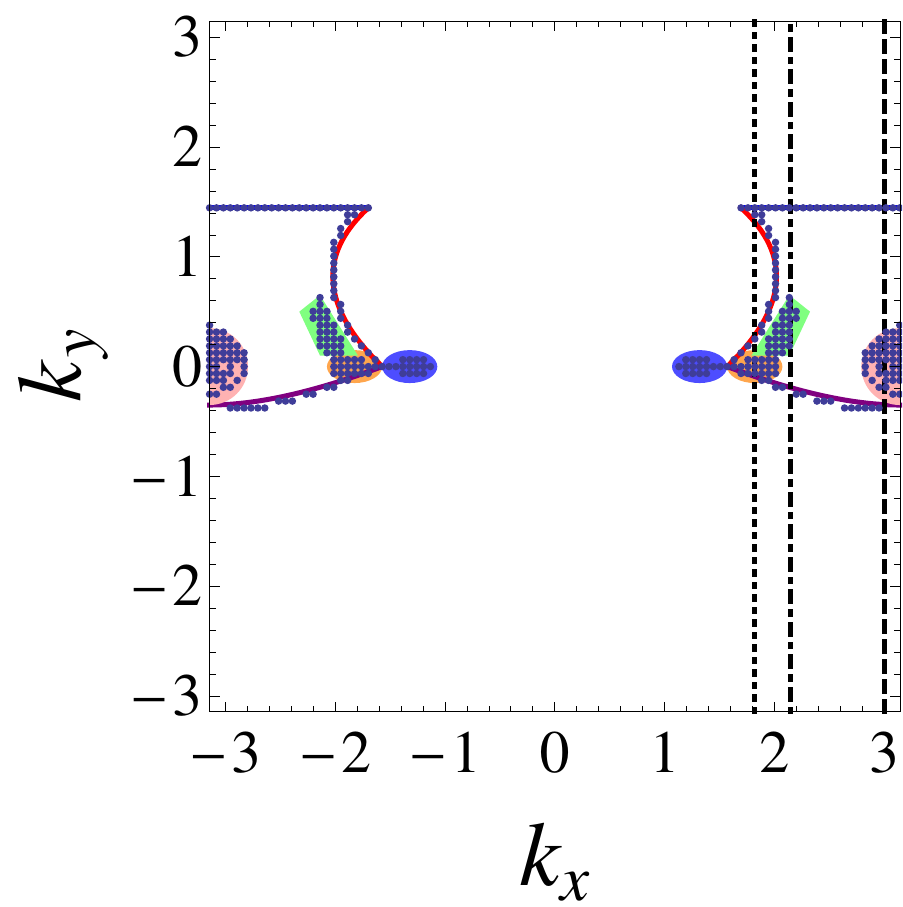}}% {figsm1c2.eps}}
		\centerline{(d)}
	\end{minipage}	
	\caption{The band structures of the Hamiltonian (17) in the main text with $L_I=L_{II}=L=100$. The first panel is for $\gamma_I=\gamma_{II}=\gamma_{TL}=1$ with $k_x=1.8,2.15,3$ (see the vertical dashed lines in the third panel). The second panel is for $\gamma_I=\gamma_{II}=1$ and $\gamma_{TL}=0.7$ with $k_x=1.8,2.15,3$. The third panel is the original data images of Fig. 2 in the main text. The zero energy state projections crossing with the vertical dashed lines can be read out from the first and second panels. {The states in the red(green, blue) bands belong to the transition layer(the bulk of type-I WSM, the bulk of type-II WSM).}  
		\label{fig5}	}
\end{figure}

\section{Weyl points, and ES}

Writing the lattice Hamiltonian (\ref{eq16}) in the main text explicitly 
\begin{equation}
H^l=d_{\mu }^l\sigma ^{\mu },
\end{equation}
where
\begin{equation}
d_{\mu }^l=\left\{
\begin{array}{ll}
d_{\mu ,II}^l, &  \gamma=\gamma_{II}, {\rm for~} n_z<n_{II}, \\
d_{\mu ,II}^l+\xi (d_{\mu ,I}^l-d_{\mu ,II}^l), &  \gamma=\gamma_{TL},{\rm for~} n_{II}<n_z<n_{I},\\
d_{\mu ,I}^l, &  \gamma=\gamma_{I}, {\rm for~} n_z>n_{I},
\end{array}
\right. \label{hi}
\end{equation}
and
\begin{eqnarray}
d_{I/II,x}&=&2\tilde t(\cos k_x-\cos K^{I/II}_x) +2\tilde t(1- \cos(k_y-K^{I/II}_y))\nonumber\\
&+&\frac{\tilde t}2(\cos 3k_x-\cos 3K^{I/II}_x)+2\tilde t\gamma (1- \cos k_z),\nonumber \\
d_{I/II,y}&=&2\tilde t \sin (k_y-K^{I/II}_y),\nonumber d_{I/II,z}=-2\tilde t \gamma \sin k_z,\nonumber\\
d_{I,0}&=&0, d_{II,0}=-2\eta_{II}(\cos k_x-\cos K^{II}_x), 
\end{eqnarray}
{where $n_z$ is the lattice site index in the $z$-direction and $\xi=\frac{n_z-L_{II}}L$.} Then the Weyl points are determined by 
\begin{equation}\
d_i^l=0 ,i=x,y,z.
\end{equation}
{At Weyl points,}
\begin{equation}
\partial_i d_j^l= 0 ,{\rm if~}   i\neq j.
\end{equation}
The Weyl points in the transition layer are running as $\xi$ and their projection to the $k_x$-$k_y$ plane is depicted in Fig. \ref{wps}{(a)}. Notice that these Weyl points {do not lie on the Fermi surface because of the interlayer coupling. If the transition layer is replaced by an insulating layer with a large band gap, then there will be ordinary Fermi arcs that connect Weyl nodes with opposite chiralities in the type-I(II) WSM on the top(bottom) surface of the insulating layer. After turning on the hopping term between the type-I and type-II WSMs in the transition layer, the Fermi arcs will reconstruct. Changing from the surface Fermi arcs into the ones connecting type-I and type-II Weyl nodes with the same chirality inside the transition layer (see the red Fermi arcs in Fig. \ref{wps}(a)).}

\begin{figure}
	\begin{minipage}{0.3\linewidth}
		\centerline{\includegraphics[width=1\linewidth]{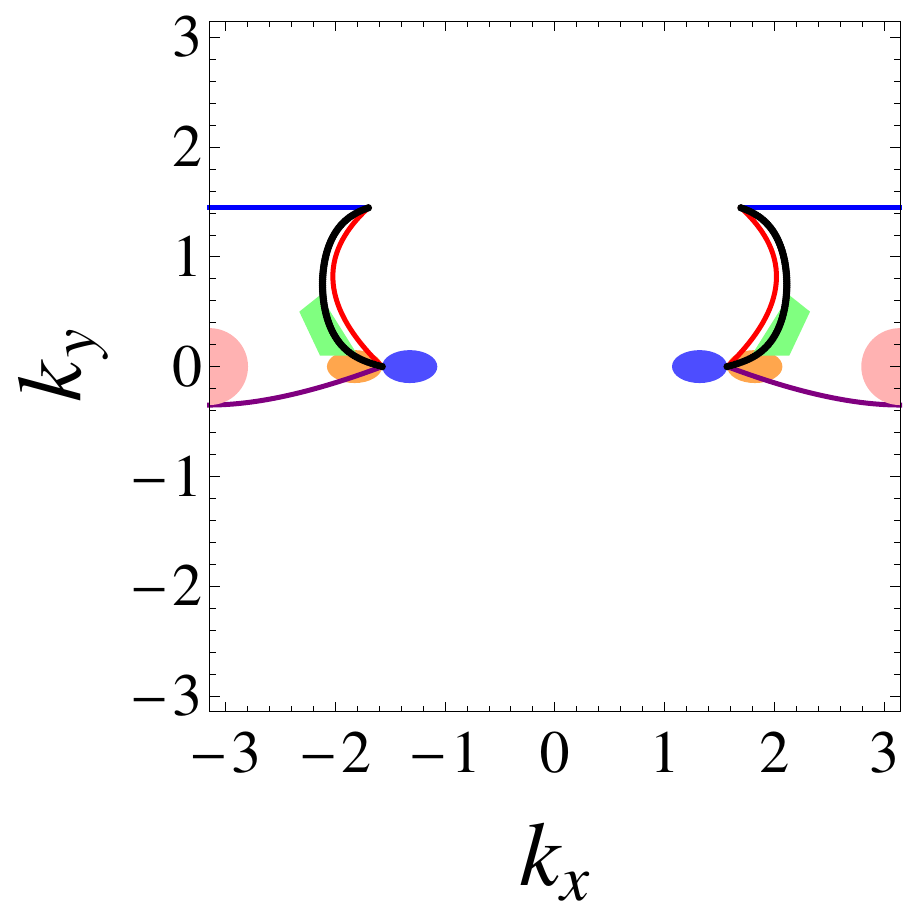}}%{Weylpoint.eps}}
		\centerline{(a)}
	\end{minipage}\vspace{4pt}
	%	\hfill
	\begin{minipage}{0.45\linewidth}
		\centerline{\includegraphics[width=1\linewidth]{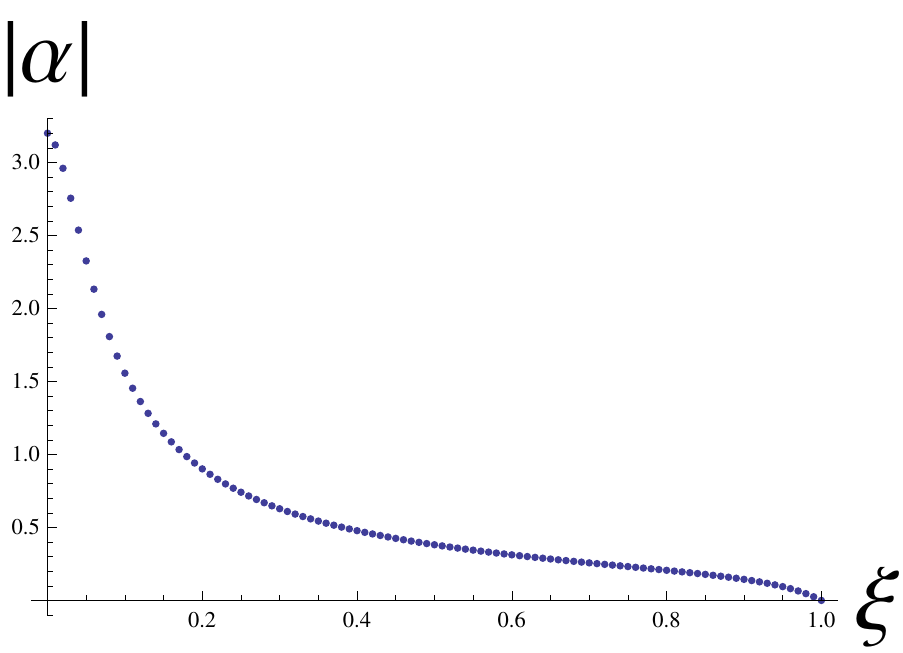}}%{alpha.eps}}
		\centerline{(b)}
	\end{minipage}
	\caption{(color online) (a) The Weyl points in the transition layer (black curves). {We copy the zero energy regions in  Fig. 2(b) of the main text for the reference.} (b) The function $|\boldsymbol{\alpha} |$ of $\xi$. $|\boldsymbol{\alpha} (\xi_h)|=1$ determines the {ES}.
		\label{wps}	}
\end{figure}

{Expanding the Hamiltonian (\ref{hi}) around the Weyl points and comparing the result with the effective Hamiltonians (3) and (4) in the main text, one obtains, 
	\begin{equation}
	|\boldsymbol{\alpha} |=\sqrt{\sum _{i=x,y,z} (\partial_i d_0^l/\partial_i d_i^l)^2}=|\partial_x d_0^l/\partial_x d_x^l|.
	\end{equation} }
The {ES} can be obtained by solving $|\boldsymbol{\alpha} |=1$ which is at $\xi_h\approx 0.18$, reading out from \ref{wps}(b).

\end{widetext}


\begin{thebibliography}{99}
\bibitem{note1} We do not discuss the case with $|\alpha_z|>1$. The {ES} is not stable in this case.
\bibitem{note} The quantum Lyapunov exponent $\lambda_{QL}$  is not necessary to be exactly the same as $\lambda_L$ due to the quantum correction. % $\lambda_L$ in the effective model increases first in $L$ increases and then decreases for the large $L$. We have yet to see such a decrease for $\lambda_{QL}$.
\end{thebibliography}
\end{document}